\documentstyle[aps,prb]{revtex}
\begin{document}
\title{Comment on ``On loss of quantum coherence in an interacting Fermi gas''
(cond-mat/9810297)}
\author{Dmitrii S. Golubev$^{1,3}$ and Andrei D. Zaikin$^{2,3}$}
\address{$^{1}$ Physics Department, Chalmers University of Technology,
S-41296 G\"oteborg, Sweden\\
$^2$ Institut f\"{u}r Theoretische Festk\"orperphysik,
Universit\"at Karlsruhe, 76128 Karlsruhe, Germany\\
$^3$ I.E.Tamm Department of Theoretical Physics, P.N.Lebedev
Physics Institute, Leninskii pr. 53, 117924 Moscow, Russia}

\maketitle

\begin{abstract}
In a recent paper \cite{EH} Eriksen and Hedegard (EH) claimed that they
``clarified'' our path integral calculation \cite{GZ1} and found an error
in our analysis. Here we will demonstrate that this claim is based on
mathematically incorrect and physically meaningless procedure invented by
the authors \cite{EH}. The main problem in the EH analysis is due
to their erroneous calculation of Gaussian path integrals and has
nothing to do with the momentum recoil of a particle interacting
with an external bath.
\end{abstract}
\vspace{0.5cm}

In a very recent preprint \cite{EH} Eriksen and Hedegard (EH) attempted to
reanalyze a part of our calculation \cite{GZ1} and ``demonstrated'' that
their version of the effective action for an electron interacting with
the rest of the fluctuating electronic environment is different
from one found in our paper \cite{GZ1}. This difference -- according to EH --
should explain disagreement of our results with ones found by Aleiner, Altshuler
and Gershenzon (AAG) \cite{AAG} and invalidate our analysis.

On one hand, we welcome an attempt of the authors \cite{EH} to go deeper
into our path integral analysis. As a result of that our work was upgraded
from ``profoundly incorrect calculation'' \cite{AAG} to being ``not without
merits'' \cite{EH}. In some sense EH also defended our work indicating that
the previous ``vague statements'' \cite{AAG} concerning ``problems in
semiclassical calculations when classical orbits intersect'' have
little to do with our calculation. We fully agree with that.

On the other hand, the results \cite{GZ1} and \cite{AAG} apparently
differ at low temperatures and according to both AAG and EH the problem
should be found in our calculation \cite{GZ1}. This is an {\it
erroneous opinion}.
In \cite{GZ2} we will demonstrate that the results of our path integral
analysis \cite{GZ1} are {\it fully equivalent} to those of AAG
at the level of the Golden-rule-type perturbative analysis employed by
the authors \cite{AAG}. In other words, if we expand our path integral
in powers of the effective action keeping only the first term of this
expansion we arrive exactly at the AAG results \cite{AAG}. The actual
difference between \cite{GZ1} and \cite{AAG} is that we {\it do not}
expand our path integral and go further than the
Golden rule approach \cite{AAG}. This issue will be discussed in details
in \cite{GZ2}.

Here, however, we are concerned with a purely technical problem.
Namely we will analyze
the claim \cite{EH} concerning the error in our path integral analysis
\cite{GZ1}. EH proceeded quite far in checking our calculations and arrived
at our Eqs. (43) \cite{GZ1} (equivalently Eqs. (11) in \cite{EH}) where
-- according to them -- the problem occurs. Although EH did not point
out where a mistake can appear in our analysis they presented their
own derivation which -- according to them -- should ``repair''
our calculation. Already at this point we would like to note that the
part of our derivation which EH would like to repair is the {\it exact
quantum mechanical calculation} involving no approximations. It yields the
effective action which -- as it was already pointed out -- on a perturbative
level allows to obtain the  results identical to those of AAG \cite{AAG}.
But it is namely {\it this} effective action which is challenged by EH.

Since we cannot exclude that EH simply misunderstood the corresponding
part of our derivation \cite{GZ1} we will reproduce it again carefully
explaining some of our steps in order to avoid possible further
misunderstandings. Then we analyze the derivation suggested by EH.

The first point where EH do not agree with our results is the path integral
reperesentation of the electron evolution operators
\begin{equation}
U_{1,2}(t_1,t_2)={\rm\bf T}\exp\left[-i\int\limits_{t_1}^{t_2}
dt'\medskip H_{1,2}(t')\right],
\label{Uj}
\end{equation}
where
\begin{eqnarray}
H_1(\bbox{p,r})=\frac{\bbox{p}^2}{2m}+U(\bbox{r}) - eV^+(t,\bbox{r})
-\frac{1}{2}\big[1-2n(H_0(\bbox{p},\bbox{r}))\big]eV^-(t,\bbox{r}),
\nonumber \\
H_2(\bbox{p,r})=\frac{\bbox{p}^2}{2m}+U(\bbox{r}) - eV^+(t,\bbox{r})
+\frac{1}{2}eV^-(t,\bbox{r})\big[1-2n(H_0(\bbox{p},\bbox{r}))\big],
\label{H12}
\end{eqnarray}
EH correctly pointed out that the operators $n(H_0(\bbox{p},\bbox{r}))$
and $V^-(t,\bbox{r})$ do not commute and this makes them different from
the ordinary single particle Hamiltonians.
However, the path integral
representation of the evolution operators (\ref{Uj}) {\it always} has the
following standard form (see e.g. \cite{Sakita}):
\begin{equation}
U_{1,2}(t_1,t_2;\bbox{r}_f,\bbox{r}_i)=
\int\limits_{\bbox{r}(t_1)=\bbox{r}_i}^{\bbox{r}(t_2)=
\bbox{r}_f}{\cal D}\bbox{r}(t')\int{\cal D}\bbox{p}(t')
\;
\exp\left[i\int\limits_{t_1}^{t_2} dt'\big( \bbox{p\dot r} -
H_{1,2}(\bbox{p},\bbox{r})\big)\right].
\label{path}
\end{equation}
The commutation rules need to be specified separately. E.g. considering the
expression for the operator $U_1$,
one can introduce an infinitesimal time shift ($+0$) and to write
the product $(1-2n)V^-$ in the action as
$[1-2n(\bbox{p}(t+0),\bbox{r}(t))]V^-(\bbox{r}(t))$. This infinitesimal
time shift guarantees that in the perturbation expansion the operator
$1-2\rho$ will be always placed before the operator $V^-$. The
expression (\ref{path}) together with the commutation rules is the
{\it exact} quantum mechanical formula.

The next step of our analysis is to calculate the kernel of the
electron evolution operator $J$ in the
presence of interaction with the effective electromagnetic environment.
This function can be expressed in terms of the path integral
\begin{eqnarray}
J(t,t';\bbox{r}_{1f},\bbox{r}_{2f};\bbox{r}_{1i},\bbox{r}_{2i})&=&
\int\limits_{\bbox{r}_1(t')=\bbox{r}_{1i}}^{\bbox{r}_1(t)=\bbox{r}_{1f}}
{\cal D}\bbox{r}_1
\int\limits_{\bbox{r}_2(t')=\bbox{r}_{2i}}^{\bbox{r}_2(t)=\bbox{r}_{2f}}
{\cal D}\bbox{r}_2\int{\cal D}\bbox{p}_1{\cal D}\bbox{p}_2\times
\nonumber \\
&&
\left\langle e^{iS_0[\bbox{r}_1,\bbox{p}_1]-iS_0[\bbox{r}_2,\bbox{p}_2]+
i\int_{t'}^tdt''\int d\bbox{r}
\big(f^-V^+ + f^+V^-\big)}\right\rangle_{V^+,V^-},
\label{J11}
\end{eqnarray}
where the electron action $S_0[x,p]$ has the form
\begin{equation}
S_0[\bbox{r},\bbox{p}]=\int\limits_{t'}^t dt''
\bigg(\bbox{p\dot r} - \frac{\bbox{p}^2}{2m} - U(\bbox{r})\bigg),
\label{S_0}
\end{equation}
and the ``charge densities'' $f^-,f^+$ are defined by the equations:
\begin{eqnarray}
f^-(t,\bbox{r})&=&e\delta (\bbox{r}-\bbox{r}_1(t))-e\delta
(\bbox{r}-\bbox{r}_2(t)),
\nonumber \\
f^+(t,\bbox{r})&=&\frac{1}{2}
\bigg(e\big[1-2n(\bbox{p}_1(t+0),\bbox{r}_1(t))\big]\delta
(\bbox{r}-\bbox{r}_1(t))+
e\big[1-2n(\bbox{p}_2(t+0),\bbox{r}_2(t))\big]
\delta (\bbox{r}-\bbox{r}_2(t))\bigg).
\label{f}
\end{eqnarray}

Averaging over the fluctuating fields $V^+,V^-$ in Eq. (\ref{J11}) amounts to
calculating Gaussian path integrals with the action
\begin{eqnarray}
iS[V^{\pm}]=
i\int\frac{d\omega d^3k}{(2\pi)^4}
V^-(-\omega,-k)\frac{k^2\epsilon(\omega,k)}{4\pi}V^+(\omega,k)-
\nonumber \\
-\frac{1}{2}
\int\frac{d\omega d^3k}{(2\pi)^4}
V^-(-\omega,-k)\frac{k^2{\rm Im}\epsilon(\omega,k)}{4\pi}
\coth\left(\frac{\omega}{2T}\right)
V^-(\omega,k);
\label{actionf}
\end{eqnarray}
where $\epsilon (\omega , k)$ is the dielectric susceptibility of the system.
This averaging can be easily (and, of course, exactly!) performed and yields
\begin{eqnarray}
J(t,t';\bbox{r}_{1f},\bbox{r}_{2f};\bbox{r}_{1i},\bbox{r}_{2i})&=&
\int\limits_{\bbox{r}_1(t')=\bbox{r}_{1i}}^{\bbox{r}_1(t)=\bbox{r}_{1f}}
{\cal D}\bbox{r}_1\int\limits_{\bbox{r}_2(t')=\bbox{r}_{2i}}^{\bbox{r}_2(t)
=\bbox{r}_{2f}}{\cal D}\bbox{r}_2\int{\cal D}\bbox{p}_1{\cal D}
\bbox{p}_2\times
\nonumber \\
&&
\times \exp\big\{iS_0[\bbox{r}_1,\bbox{p}_1]-iS_0[\bbox{r}_2,\bbox{p}_2]-
iS_R[\bbox{r}_1,\bbox{p}_1,\bbox{r}_2,\bbox{p}_2]-
S_I[\bbox{r}_1,\bbox{r}_2]\big\};
\label{J}
\end{eqnarray}
where
\begin{eqnarray}
S_R[\bbox{r}_1,\bbox{p}_1,\bbox{r}_2,\bbox{p}_2]&=&
\frac{e^2}{2}\int\limits_{t'}^t dt_1 \int\limits_{t'}^t dt_2
\big\{R(t_1-t_2,\bbox{r}_1(t_1)-\bbox{r}_1(t_2))
\big[1-2n\big(\bbox{p}_1(t_2+0),\bbox{r}_1(t_2)\big)\big]-
\nonumber \\
&&
-R(t_1-t_2,\bbox{r}_2(t_1)-\bbox{r}_2(t_2))
\big[1-2n\big(\bbox{p}_2(t_2+0),\bbox{r}_2(t_2)\big)\big]
\nonumber \\
&&
+R(t_1-t_2,\bbox{r}_1(t_1)-\bbox{r}_2(t_2))
\big[1-2n\big(\bbox{p}_2(t_2+0),\bbox{r}_2(t_2)\big)\big]-
\nonumber \\
&&
-R(t_1-t_2,\bbox{r}_2(t_1)-\bbox{r}_1(t_2))
\big[1-2n\big(\bbox{p}_1(t_2+0),\bbox{r}_1(t_2)\big)\big]
\big\};
\label{SR}
\end{eqnarray}
and
\begin{eqnarray}
S_I[\bbox{r}_1,\bbox{r}_2]&=&
\frac{e^2}{2}\int\limits_{t'}^t dt_1 \int\limits_{t'}^t dt_2
\bigg\{I(t_1-t_2,\bbox{r}_1(t_1)-\bbox{r}_1(t_2))+
I(t_1-t_2,\bbox{r}_2(t_1)-\bbox{r}_2(t_2))-
\nonumber \\
&&
-I(t_1-t_2,\bbox{r}_1(t_1)-\bbox{r}_2(t_2))-
I(t_1-t_2,\bbox{r}_2(t_1)-\bbox{r}_1(t_2))\bigg\}.
\label{SI}
\end{eqnarray}
The functions $R$ and $I$ are defined by the equations
\begin{eqnarray}
R(t,\bbox{r})&=&\int\frac{d\omega d^3k}{(2\pi)^4}\medskip
\frac{4\pi}{k^2\epsilon(\omega,k)}e^{-i\omega t+i\bbox{kr}}
\label{R}\\
I(t,\bbox{r})&=&\int\frac{d\omega d^3k}{(2\pi)^4}\medskip
{\rm Im}\left(\frac{-4\pi}{k^2\epsilon(\omega,k)}\right)
\coth\bigg(\frac{\omega}{2T}\bigg)
e^{-i\omega t+i\bbox{kr}} .
\label{RI}
\end{eqnarray}

The combination in the exponent (\ref{J}) is the effective action. It is
defined on the Keldysh contour, the last two terms $S_R$ and $S_I$
depend on two coordinates $\bbox{r}_1$ and $\bbox{r}_2$ and are defined
by nonlocal in time expressions (\ref{SR}, \ref{SI}).
The influence of the environment,
the Pauli principle and the commutation rules between $\rho$ and $V^-$ are
{\it fully} accounted for. Proceeding from (\ref{Uj}) to (\ref{RI}) we made
{\it no approximations}. The quasiclassical approximation is used already
at a later stage of our calculation. Then the infinitesimal time shift
does not play any role and can be ignored. But since EH challenge
our effective action, namely its part (\ref{SR}), we will not discuss
further steps of our calculation \cite{GZ1}.

Let us now examine the derivation suggested by EH. They define the quantity
\begin{equation}
I=\langle p_f, x_f|U_1(t)|x_i, p_i \rangle _{V^{\pm}}
\label{I}
\end{equation}
(their Eq. (20)) which -- although can be considered as a mathematical object --
has little physical meaning in our problem. Note that $U_1$ here is
the evolution operator defined only on the forward part of the Keldysh
contour, while both the fields $V^{\pm}$ and the effective action are
(and should be!) defined on both forward and backward part of this contour.
Already here EH have a problem: how is it possible to derive the
effective action in the presence of the environment working
only with one part of the Keldysh contour? This contradics to all
what is known about this matter. But this is a textbook issue (see e.g.
\cite{FH}) and we are not going to discuss it in more detais here.

Since EH would like to calculate the quantity $I$ (\ref{I}) we will first
demonstrate how to do it correctly. If one uses the exact quantum
path integral representation (\ref{Uj}) of the
operator $U_1(t)$, its kernel can be defined as
\begin{equation}
J_1(t,0;\bbox{x}_{f},\bbox{x}_{i})=
\int\limits_{\bbox{r}(0)=\bbox{x}_{i}}^{\bbox{r}(t)=\bbox{x}_{f}}
{\cal D}\bbox{r}\int{\cal D}\bbox{p}
e^{iS_0[\bbox{r},\bbox{p}]+
i\int_{0}^t dt\int d\bbox{r}
\big(f_1^-V^+ + f_1^+V^-\big)},
\label{J12}
\end{equation}
with
\begin{eqnarray}
f_1^-(t,\bbox{r})&=&e\delta (\bbox{r}-\bbox{r}(t)),
\nonumber \\
f_1^+(t,\bbox{r})&=&(e/2)\big[1-2n(\bbox{p}(t+0),\bbox{r}(t))\big]\delta
(\bbox{r}-\bbox{r}(t)).
\label{f1}
\end{eqnarray}
The equations (\ref{J12}) and (\ref{f1}) should now be combined with (\ref{I}).
The Gaussian path integral over the fields $V^+$ and $V^-$ can be evaluated
exactly and one finds
\begin{equation}
I=
\int\limits_{\bbox{r}(0)=\bbox{x}_{i}}^{\bbox{r}(t)=\bbox{x}_{f}}
{\cal D}\bbox{r}\int{\cal D}\bbox{p}
e^{iS_0[\bbox{r},\bbox{p}]+iS_{eff}[\bbox{r},\bbox{p}]},
\label{I1}
\end{equation}
where the action $S_{eff}$ is given by
\begin{eqnarray}
S_{eff}&=&
e^2\int\limits_0^t dt' \int\limits_0^{t'} dt''
\bigg(i\big\langle V^+(t',\bbox{r}(t'))V^+(t'',\bbox{r}(t''))\big\rangle
\nonumber \\
&&
+\frac{1}{2}[1-2n(\bbox{p}(t''+0),\bbox{r}(t''))]
\big\langle V^+(t',\bbox{r}(t'))V^-(t'',\bbox{r}(t''))\big\rangle\bigg),
\label{Seff}
\end{eqnarray}
This action depends on both coordinate $\bbox{r}$ and momentum $\bbox{p}$.
It does not contain any recoil! We repeat: the quantities $I$ (\ref{I}) and
$S_{eff}$ (\ref{Seff}) by itself have little physical meaning. But formally they
can be defined and calculated as it was shown above.

What do EH suggest instead of this simple exact calculation?

To make the comparison with EH results easier, it
is useful to rewrite the action (\ref{Seff}) in the following form:
\begin{equation}
S_{eff}[\bbox{r},\bbox{p}]=\frac{i}{2}
\left\langle\left(
\int\limits_0^t dt\int d\bbox{r}\left( f_1^-V^+ + f_1^+V^-\right)
\right)^2\right\rangle_{V^\pm}
\label{Sef}
\end{equation}

In order to evaluate their quantity $I$ (\ref{I}) EH used the procedure which is
nothing else but the expansion of the exact expression (\ref{J12}) in powers
of $V^\pm$ up to the second order, averaging this first order correction over
the fluctuating potentials and writing it again in the form of the exponent.
Namely, their procedure reduces to the following set of approximations
\begin{eqnarray}
I&=&\left\langle
e^{iS_0[\bbox{r},\bbox{p}]+
i\int_{0}^t dt\int d\bbox{r}
\big(f_1^-V^+ + f_1^+V^-\big)}\right\rangle_{V^\pm,\bbox{p},\bbox{r}}
\nonumber\\
&\simeq&
\left\langle
e^{iS_0[\bbox{r},\bbox{p}]}\left[1-\frac{1}{2}
\left(\int_{0}^t dt
\big(f_1^-V^+ f_1^+V^-\big)\right)^2\right]
\right\rangle_{V^\pm,\bbox{p},\bbox{r}}
\end{eqnarray}
Then they calculate the first order correction given by the last term in
the square brackets and call the following construction ``effective action'':
\begin{equation}
S_{eff}^{EH}[\bbox{r}]=\frac{i}{2} \frac{\int{\cal
D}\bbox{p}\;\; e^{iS_0[\bbox{r},\bbox{p}]} \left\langle\left( \int\limits_0^t
dt\int d\bbox{r}\left( f_1^-V^+ + f_1^+V^-\right) \right)^2\right\rangle_{V^\pm}}
{\int{\cal D}\bbox{p}\;\; e^{iS_0[\bbox{r},\bbox{p}]}}
\label{SEH}
\end{equation}
This ``action'' is very different from the exact result
(\ref{Seff},\ref{Sef}) and is just slightly modified first order of the
perturbation
theory in interaction. EH do not even bother calculating
this ``action'' and leave it in the form equivalent to
(\ref{SEH}) (see their eq. (27)). They evaluate it only on a straight line
trajectory (their eqs. (28,29)). Apart from the fact that these trajectories
are meaningless in the case of disordered metals at all relevant distances
(i.e. distances longer than the mean free path), we do not understand how
one can compare the action as a {\it functional} of all trajectories
with a {\it number} calculated for one trajectory only.

Anyway, in contrast to the EH claim the problem has nothing to do
with the commutation relations but rather emerges from a wrong calculation
suggested by EH. In order to illustrate that one can consider
e.g. the muon motion which is not restricted by the Pauli
principle. In this case the terms containing $n(H_0(\bbox{p},\bbox{r}))$
in the Hamiltonians (\ref{H12}) are absent and the problems with commutation of
the operators do not appear at all. However, even in that case the exact action
(\ref{Seff},\ref{Sef}) remains very different from the action (\ref{SEH}).
This example
demonstrates that the procedure used
by the authors \cite{EH} fails even in the simplest case.

Now let us discuss the momentum recoil. EH concluded that our theory
does not describe the effect of recoil because it is not observed in the action.
This is wrong. Our expression for the effective action was obtained by the
{\it exact} (within RPA) integration over all the degrees of freedom of the
effective environment and hence includes all processes in all orders.
It would be therefore very naive to try to interpret the full action only in
terms of the simplest first order scattering processes. However, if one
wants to describe only such processes one can easily do that
proceeding perturbatively and expanding the path integral
in powers of the exact action (\ref{J}-\ref{SI}).
All correct recoils and other effects are fully reproduced.
It was demostrated e.g. in Sec. 5 of our paper \cite{GZ1}.
There EH can easily find e.g. the combination $n_{p-k}$ entering eqs.
(90), (91), (93-95) or ``the standard result from the standard many
body theory'' containing the combination ``coth-tanh'' (see eq. (97) of
\cite{GZ1}). But all that is only the simple Golden rule perturbation
theory which is actually known to fail at low $T$. It is really hard
to understand how the exact procedure can be ``repaired'' by making
a rough approximation!

One can assume that if one proceeds perturbatively in the interaction
the expansion done by EH is legitimate. This is not true: although
interaction may be small, the time integrals are
large, and the corresponding terms should be always kept in the exponent in the
long time limit. If one wants to expand one should keep {\it all} orders
of this expansion. But to be on a safe side it is always better to calculate
exactly as long as one can do it. This will never make things worse.

Just for an illustration let us make a trivial exercise. Consider a Gaussian
integral
\begin{equation}
I_1(a,b)=\int dx \exp (-ax^2-2bx)
\end{equation}
In order to find out what is the leading dependence of $I_1$ on $b$ it is not
even necessary to calculate this trivial integral. One just makes a shift and
obtains $I_1 \propto \exp (b^2/a)$. Then one can decide if one wants to
expand the exponent or not. However if one expands in $b$ first the
result may be wrong depending on the value $b^2/a$. This seems to be
obvious. Consider now a slightly more complicated Gaussian integral
\begin{equation}
I_N[a,b]=\int\prod_{n=1}^{N} dx_n \exp (-a_nx_n^2-2b_nx_n)
\end{equation}
Analogously one obtains
\begin{equation}
I_N \propto \exp (\sum_{n=1}^N b^2_n/a_n)
\end{equation}
Again if the {\it whole sum} in the exponent is small one can expand. However,
if one expands in all $b_n$ first and calculates the integrals over all $x_n$
one takes a risk of making a mistake. Some of $b^2_n/a_n$ may be small, some
others not and -- especially if $N$ is very large -- the sum in the exponent
will not be small.

EH's -- and not only EH's -- critique of our paper \cite{GZ1} is essentially
on this level: ``expand first, get ``coth-tanh'', obtain zero result
at $T \to 0$, compare with our result \cite{GZ1}, observe the difference,
conclude that our result \cite{GZ1} is wrong''. But we agree (and always
knew it) that one gets ``coth-tanh'' if one expands first.
What we are saying is that {\it it is wrong to expand first}. We welcome
constructive critique of our results. But it is always better to understand
{\it what} was actually done in the paper and criticize afterwards
than vise versa.

Summarizing, the EH claim that we made an error ``when dealing with the
real space path integral representation of quasiparticle propagation'' is
wrong. [By the way, we do not quite know what the quasiparticle in a
disordered metal with interaction is, do EH know?] In contrast to EH
beliefs our formalism does ``allow for tiny changes in the momentum
and energy''. It actually does it in all orders and not only in the first
order like in \cite{EH}. The erroneous result indeed occurs, it occurs
in the EH paper \cite{EH} and not because of ``tiny changes'' but rather
because the authors \cite{EH} believe that it is more accurate to
calculate Gaussian integrals perturbatively than to do it exactly.
The story with the EH paper \cite{EH} is over.

\end{document}